\documentclass[%
 reprint,
 amsmath,amssymb,
 aps,
]{revtex4-2}
\usepackage{graphicx}
\usepackage{dcolumn}
\usepackage{bm}
\usepackage[colorlinks=true,citecolor=red,linkcolor=blue]{hyperref}
\usepackage[mathlines]{lineno}
\usepackage[capitalize]{cleveref}
\usepackage{subfigure}

\def\Tr{{\rm{Tr}}}
\def\Re{{\rm{Re}}}
\def\Im{{\rm{Im}}}

\begin{document}

\preprint{APS/123-QED}

\title{Quantum Information Scrambling and Entanglement in Bipartite Quantum States}

\author{Kapil K. Sharma}
 \altaffiliation[Former Affiliation: LIT, JINR Dubna \\]{iitbkapil@gmail.com}
\author{Vladimir P. Gerdt}%
 \email{gerdt@jinr.ru}
\affiliation{$^{\star}$DY Patil International University,
Sect-29, Nigdi Pradhikaran, Akurdi,
Pune, Maharashtra-411044, India \\ \\
$^{\dagger}$Joint Institute for Nuclear Research, \\ 6 Joliot-Curie St, 141980 Dubna, Russia \\
and \\
Peoples' Friendship University of Russia (RUDN University) \\
6 Miklukho-Maklaya St, 117198 Moscow, Russia
}
\date{\today}

\begin{abstract}
Investigating the influence of quantum information (QI) scrambling on quantum correlations in a physical system is an interesting problem. In this article we establish the mathematical connections among the quantifiers known as quantum information scrambling, Uhlmann fidelity, Bures metric and bipartite concurrence. We study these connections via four point out-of-time-order correlation (OTOC) function used for quantum information scrambling. Further we study the dynamics of all the quantifiers and investigate the influence of QI scrambling on entanglement in two qubits prepared in Bell states. We also determine the related QI scrambling and entanglement balancing points and investigate that they are periodic for the Ising Hamiltonian.

\end{abstract}

\maketitle\section{Introduction}
Quantum information scrambling (QI scrambling)\,\citep{qi1,qi2,qi3,vqi} is known as the spreading of the quantum information over the physical system. The phenomenon of quantum information scrambling may takes place in any physical system because of the chaotic situations. From classical mechanics perspective, the chaos is observed by studying the dynamics of classical trajectories in phase space. If the initial condition is very sensitive, then the trajectories diverge in the space and follow the Lyapunov exponent as $e^{\lambda t}$\,\cite{lay1,lay2}. This diversion of trajectories in the phase space is known as butterfly effect\,\cite{bf1}. It is difficult to study butterfly effect in quantum mechanics because a notion of trajectories is missing. However, the notion of trajectories takes place not only in classical but also in semi-classical domain\,\cite{scq1}.

The chaos in the quantum domain does not have a single definition, and there are many ways to look into quantum chaos from the quantum mechanics point of view\,\cite{qc1}. However the notion of trajectories takes place in classical or semi-classical domain\,\cite{scq1}.
A popular approach to study the chaos in the physical system is to measure the degree of irreversibility by using the mismatch between forward-backward evolution of the system. In order to quantify the chaos by using forward-backward evolution approach, the famous quantifiers are known as Loschdimdt Echo and irreversible entropy production\,\cite{los1,los2,ep1}. These quantifiers also have experimental manifestations in varieties of physical systems. In connection of forward-backward evolution approach, recently out-of-time-order correlators (OTOC) attracted much attention to measure the quantum information scrambling in thermal density matrices\,\cite{ot2}. OTOC was originally discovered by Larkin and Ovchinnikov in their application of the quasi-classical method to theory of superconductivity in 1968\,\cite{ot1}. They studied the behavior of classical pair correlator function i.e. $C_{c}(t)=\langle [p(t)p(0)]^{\dagger}.[p(t)p(0)]\rangle=e^{2\lambda t}$ in Fermi gas with the Lyapunov exponent $\lambda$. The Lyapunov exponent quantifies the strength of the chaos which is unbounded for classical physical systems while it has a bound for quantum systems with the limit $\lambda\leq 2\pi KT/\hbar$. Recent trends in quantum chaos deal with the quantum mechanical version of $C_{c}(t)$, represented in terms of quantum operators. The quantum version of OTOC has close connection with the black hole information problem and with holographic theories of quantum chaos in many body physics. It has been shown by Hayden and Preskill\,\cite{qi1} by considering a simple model of random unitary evolution that black holes rapidly process the quantum information and exhibit the fastest information scrambling. Temperature is natural recourse of energy for black holes, which is a major factor for information scrambling; hence it is customary to study the quantum information scrambling in thermal density matrices. Continuing the OTOC discussion, here we mention that the investigation of different versions of OTOC is also an active area. The impact of OTOC can disturb the quantum correlations in a physical system but simultaneously the deep distinction between quantum information scrambling and decoherence is not clear\,\cite{qid1}.

A lot of work have been carried out on OTOC in different physical systems dealing with varieties of domains like conformal field theories, quantum phase transition, Luttinger liquids, quantum Ising chain, symmetric Kitaev chain, quadratic fermions, hardcore boson model, XX spin chain with random field\,
\cite{cf1,lt1,lt2,qch1,fot1,xot1}. Often, the OTOC in spin chains have been studied as a function of the distance between two arbitrary spins which are imposed by actions of the local non-commutative operators\,\cite{scv1}. Further the Lyapunov exponent as a function of the velocity, i.e. $\lambda(v)$, has been studied in classical and semiclassical regime and early time behavior of quantum information scrambling has been investigated with Baker-Campbell-Hausdorff (BCH) formula\,\cite{bch1,bch2}. We recall that at present the quantum chaos community is bound to study OTOC in thermal density matrices by following the analogy of the involvement of the temperature with black holes. However, dynamical studies in varieties of non-thermal quantum states are missing.

In the given work, we establish the direct mathematical connection between quantum information scrambling and concurrence in pure bipartite quantum states\,\cite{con1,con2,con3}. With this mathematical connection, it is easy to study the direct influence of quantum information scrambling on the entanglement. However, establishing such direct connections is a hard problem for larger Hilbert spaces. We also establish mathematical connections between quantum information scrambling, Uhlmann fidelity and Bures metric\,\cite{uh1,bm1}. These quantifiers are helpful to study the degree of mismatch between forward and backward evolution and the influence of this mismatch on QI scrambling. Further we study the dynamical behavior of the above mentioned quantifiers in two-qubit Bell states\,\cite{bl1}, which proves the strength of mathematical relations.

To start with, the quantum mechanical version of OTOC is given by
the expectation value of the operator
\begin{equation}
C(t) = [W(t),V]^{\dagger}.[W(t),V]\,. \label{e1}
\end{equation}
In quantum mechanics, the expectation value of an operator $O$ is defined as
$\langle \,O\rangle_\rho={\Tr}[O.\rho\,]$,
where $O$ is the operator and $\rho$ represents the density matrix of a quantum state.

The expectation value of $C(t)$ is represented in Heisenberg picture, and we assume that the operators  $W(t)$ and $V$ are Hermitian as well as unitary.
For initial stage at $t=0$, the operators $\{W(0),V\}$ commute and no QI scrambling takes place. It is expressed by  the condition
\[
[W(0),V]=0\,.
\]
As time advances, the commutativity between $W(t)$ and $V$ may break, which produce QI scrambling. So the condition for the existence of QI scrambling can be considered as
\begin{equation}
[W(t),V]\neq 0\,. \label{Noncommute}
\end{equation}
The unitary time evolution of the operator $W(t)$ under a certain Hamiltonian governs the degree of commutativity and hence information scrambling. The unitary time evolution of the operator $W(t)$ is given by the series
\begin{eqnarray}
&& W(t)=e^{iHt}W(0)e^{-iHt}=W(0)+it[H,W(0)]+ \nonumber \\
&& \frac{t^{2}}{2!}[H,[H,W(0)]]+\frac{it^{3}}{3!}[H,[H,[H,W(0)]]]+\ldots \label{se1}
\end{eqnarray}
At $t=0$, the series yields $W(0)$ and no scrambling takes place. For the existence of QI scrambling, the following condition should also be satisfied,
$[H,W(0)]\neq 0$. Here we mention that if $\{H,W(0)\}$ are bounded operators with $||H||\leq \epsilon$ and $||W(0)||\leq \epsilon$, then the series is a convergent and it is helpful to study the behavior of QI scrambling.

\section{Linking QI scrambing, Uhlmann Fedility and Bures metric} \label{a1}
In this section we explore the derivation of the OTOC and establish the mathematical connections among the quantifiers known as QI scrambling, Uhlmann fidelity and Bures metric for pure quantum states. We recall the definition of OTOC given in Eq.\eqref{e1} as
\begin{equation}
C(t)=[W(t),V]^{\dagger}.[W(t),V]=P.Q \label{me}
\end{equation}
where
\[
P=[W(t),V]^{\dagger}=V^{\dagger}.W(t)^{\dagger}-W(t)^{\dagger}.V^{\dagger}
\]
and
\begin{equation*}
Q=[W(t),V]=W(t).V-V.W(t)\,.
\end{equation*}
The product $P.Q$ reads
\begin{eqnarray}
P.Q=V^{\dagger}.W(t)^{\dagger}.W(t).V-V^{\dagger}.W(t)^{\dagger}.V.W(t)\nonumber \\ -W(t)^{\dagger}.V^{\dagger}.W(t).V+W(t)^{\dagger}.V^{\dagger}.V.W(t)\label{eq:7}
\end{eqnarray}
Here $W(0)$ and $V$ are Hermitian operators, so they must satisfy
the following conditions,
\begin{equation}
W(0)^{\dagger}=W(0), \quad W(t)^{\dagger}=W(t) \label{o1}
\end{equation}
and
\begin{equation}
V^{\dagger}=V\,. \label{o2}
\end{equation}
Applying these conditions to Eq.\,\eqref{eq:7}, we obtain
\begin{eqnarray}
P.Q=2.I-V.W(t).V.W(t)-W(t).V.W(t).V  \nonumber \\
=2.I-\{W(t).V.W(t).V\}^\dagger-W(t).V.W(t).V \label{eq:12}
\end{eqnarray}
with the identity matrix $I$.

Taking the average on both sides with the density matrix $\rho$ and applying the cyclic property of the trace operation, our simplification yields
\[
\langle P.Q\rangle_\rho =2(1-{\Re}\{{\Tr}[W(t).V.W(t).V.\rho]\})\,.
\]

Taking into account Eqs.(\ref{me}) and (\ref{eq:12}) we obtain the expression of OTOC
\[
\langle C(t)\rangle_\rho=2\, [1-{\Re}\{Z\}]\,,
\]
where
\begin{equation}
Z={\Tr}[M]\,,\quad M=W(t).V.W(t).V.\rho\,.\label{d1}
\end{equation}
So in short we can write
\begin{equation}
F_{R}(t)={\Re}\{Z\}\,. \label{r1}
\end{equation}

Finally, the expectation value of $P.Q$ in Eq.\,\eqref{eq:12} with the factor can be written over a density matrix $\rho$ as
\begin{equation}
\langle  P.Q \rangle_\rho = \langle C(t)\rangle_\rho=2\,[1-F_{R}(t)]\,.\label{is1}
\end{equation}
We emphasize that the above equation involves $F_{R}(t)$ as expressed in Eqs.\,\eqref{d1}--\eqref{r1} over a density matrix $\rho$.
To establish the OTOC connection with the Uhlmann fidelity of pure quantum states, we reconsider Eqs.\,\eqref{d1}--\eqref{r1}  and rewrite the equation by using the cyclic property of trace operation
\[
F_R(t)={\Re}[\langle\psi|W(t).V.W(t).V|\psi\rangle
]={\Re}[\langle y|x\rangle]
\]
with
\begin{equation}
|x\rangle=W(t).V|\psi\rangle \label{ef1}
\end{equation}
and
\begin{equation}
|y\rangle=V.W(t)|\psi\rangle\,. \label{eb1}
\end{equation}

Here the state $|x\rangle$ represents forward evolution of the state $|\psi\rangle$ and $\langle y|$ represent the backward evolution of the state $|\psi\rangle$ under the actions of the operators $\{W(t),V\}$. Since the operators $W(t)$ and $V$ are unitary, the state $|\psi\rangle$ does not loose its purity during the complete evolution. Here under the complete evolution we mean the total evolution involving the forward and backward evolution. The Uhlmann fidelity of two quantum states gives information about the overlapping of quantum states and measures the similarity or probability of transition between two states. The Uhlmann fidelity between two pure quantum states
$\{|x \rangle,|y\rangle\}$ is defined as
\begin{equation}
f=|\langle y|x\rangle|^{2}\,. \label{uf1}
\end{equation}
The above equation gives the degree of mismatch between the forward and backward evolution of the quantum state Eq.\eqref{d1}, and $f$ is a real quantity. The range of the fidelity is $f\in[0,1]$. We plug-in the values from Eqs.\,\eqref{ef1} and \eqref{eb1} into Eq.\,\eqref{uf1}. Then we obtain
\[
f=|Z|^{2}
\]
or taking into account Eqs.\,\eqref{d1}--\eqref{r1},
\begin{equation}
f=\Big(F_{R}(t)\Big)^{2}+\Big({\Im}\{Z\}\Big)^{2}\,,\quad \quad \sqrt{f}=\big|Z\big|\,. \label{f1}
\end{equation}
Here the density matrix $\rho=|\psi \rangle \langle \psi |$ (cf.~\eqref{d1}) is used in the context of pure quantum states.

\noindent
Given a mixed (not pure) state in Eq.~\eqref{d1}
\[
\rho=\sum_{i=1}^np_i|\psi_i\rangle\langle\psi_i|\,,\quad n\geq 2\,, \quad p_i\geq 0\,,\quad  \sum_{i=1}^np_i=1\,,
\]
defined on a Hilbert space $\mathcal{H}_A$, one can take a purification~(cf.\,\cite{Kleinmann:2006}) of $\rho$ defined on a Hilbert space $\mathcal{H}_A\otimes\mathcal{H}_B$
\[
|\Psi\rangle=\sum_{i=1}^n \sqrt{p_i}|\psi_i\rangle|\phi_i\rangle\,,\quad {\Tr}_B\big(|\Psi\rangle\langle \Psi|\big)=\rho\,,
\]
where ${\Tr}_B$ denotes partial trace over the subsystem $B$. Thus, Eqs.\,\eqref{d1} can be written as
\[
   Z=\langle \Psi | \Big(W(t).V.W(t).V\otimes I_B\big)|\Psi\rangle\,,
\]
where $I_B$ is the identity matrix acting on elements in $\mathcal{H}_B$. Accordingly, instead of Eq.~\eqref{ef1} and Eq.~\eqref{eb1} we can consider the pure states
\[
   W(t).V\otimes I_B|\Psi\rangle  \quad \text{and}\quad  V.W(t)\otimes I_B |\Psi\rangle\,.
\]

From Eqs.\,\eqref{is1} and \eqref{f1} it follows
\begin{equation}\label{UpperBound}
F_{R}(t)=\sqrt{f-\Big({\Im}\{Z\}\Big)^{2}}
\end{equation}
and
\begin{equation}
\langle C(t) \rangle_\rho=2\Big[1-\sqrt{f-\Big({\Im}\{Z\}\Big)^{2}}\Big]\,. \label{es1}
\end{equation}
The Eq.(\ref{es1}) establish the connection between Uhlmann fidelity and QI scrambling. If the imaginary part is zero,
\begin{equation}
{\Im}\{Z\}=0\,, \label{ImZ:0}
\end{equation}
then $\langle C(t) \rangle_\rho$ takes the following form
\begin{equation}
\langle C(t) \rangle_\rho=2\Big[1-\sqrt{f}\Big]\,.\label{qz1}
\end{equation}
Equations~\eqref{UpperBound} and~\eqref{qz1} show that $\sqrt{f}$ is the exact (i.e.,  reachable) upper bound on $F_R(t)$ and $2[1-\sqrt{f}]$ is the exact lower bound on the QI scrambling $\langle C(t)\rangle_\rho$, respectively. The indicated bounds are reached when $Z$ becomes real. Below we investigate connection between the QI scrambling and Bures metric. Referring to~\cite{bl2}, we know that the relation between Bures metric $D$ and Uhlmann fidelity is given by
\begin{equation}
D=\sqrt{2(1-\sqrt{f})} \label{bm1}
\end{equation}
Adjusting the value of $f$ from Eq.(\ref{bm1}) into Eq.(\ref{es1}), we obtain
\begin{equation}
\langle C(t)\rangle_{\rho}=2\Big[1-\sqrt{\Big(1-\frac{D^{2}}{2}\Big)^{2}-\Big({\Im}\{Z\}\Big)^{2}}\Big]\label{na1}
\end{equation}
Under the condition~\,\eqref{ImZ:0} the Eq.\,(\ref{na1}) reads
\[
\langle C(t) \rangle_\rho=D^{2}\,.
\]

\begin{figure*}
\includegraphics[scale=1.2]{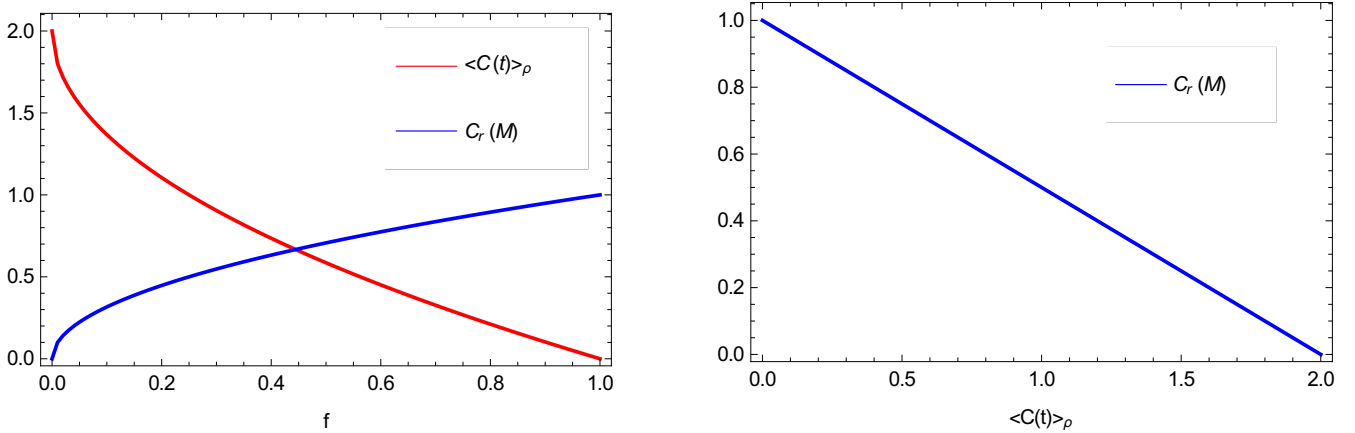}
\caption{Left Figure: Plot of QI scrambling $\langle C(t)\rangle_\rho$ and concurrence $C_{r}(M)$
vs. Uhlmann fidelity $f$. Right Figure: Plot of concurrence $C_{r}(M)$ vs. QI scrambling $\langle C(t)\rangle_\rho$.} \label{fg1}
\end{figure*}
The last equation establishes the relation between QI scrambling and Bures metric. We mention that Bures metric is another measure of closeness of two quantum states and related  with Uhlmann fidelity. Here we find that under the condition~\eqref{ImZ:0} the QI scrambling is a square function of the Bures metric. In the present paper we only focus on Uhlmann fidelity as this has lucid properties and is widely used in literature.

\subsection{Important properties of QI scrambling}
Bases on the Eq.\eqref{es1} we find the following properties of QI scrambling,
\begin{itemize}
\item[1.] Positivity, $\langle C(t) \rangle_\rho\geq 0$\,.
\item[2.] Bounded limits, $0\leq\langle C(t) \rangle_\rho \leq 2.$
\item[3.] Unitary invariant, $U C(t)\, U^{\dagger}= C(t)$\,.
\item[4.] Exchange symmetry over forward and backward evolution. The Uhlmann fidelity has the following obvious symmetry
\[
f=|\langle y|x\rangle|^{2}=|\langle x|y\rangle|^{2}\,.
\]
Hence, we conclude that under the exchange of forward and backward evolution the QI scrambling $\langle C(t)\rangle_\rho$ will remain the same.
\end{itemize}

\subsection{Linking QI scrambling and two-qubit concurrence}

In this section we establish the mathematical connection between QI scrambling and the
two-qubit concurrence. Here we mention that concurrence $C_{r}$ is a good measure of entanglement for two-qubit systems, and it has its experimental manifestations. The concurrence is defined as
\begin{equation}
C_{r}(|\psi\rangle)=|\langle\psi|\sigma_{y}\otimes\sigma_{y}|\psi^{\star}\rangle| \label{conc1}
\end{equation}
or
\begin{equation}
C_{r}(|\psi\rangle)=\big|{\Tr}[(\sigma_{y}\otimes \sigma_{y}).(|\psi^{\star}\rangle\langle\psi|)]\big|\,. \label{c11}
\end{equation}
Here $\sigma_{y} \otimes \sigma_{y}$ is the spin flip matrix, which flips both the spins under the action of Pauli Y operator, and $|\psi^{\star}\rangle$ is the complex conjugate of the state $|\psi\rangle$. If we consider the case $(|\psi^{\star}\rangle=|\psi\rangle)$, which means that the density matrix $\rho$ has only real entries, and the condition $\rho^{\star}=\rho$ is satisfied. By using this condition, we may write the Eq.\,\eqref{c11} as
\[
C_{r}(\rho)=\big|{\Tr}[(\sigma_{y}\otimes \sigma_{y}).\rho]\big|
\]
Furthermore, we may straightforwardly write the concurrence in the state $M$ given by Eq.\,\eqref{d1}
\[
C_{r}(M)=\big|{\Tr}[(\sigma_{y}\otimes \sigma_{y}).M\big|
\]
or
\begin{equation}
C_{r}(M)=\big|{\Tr}[(\sigma_{y}\otimes \sigma_{y}).W(t).V.W(t).V.\rho]\big|\,. \label{cc1}
\end{equation}

We know that the product of two Hermitian matrices $A$ and $B$ is Hermitian if and only if $[A,B]=0$. The product $W(t).V$ may not be Hermitian, in general, because of the condition $[W(t).V]\neq 0$ (cf.\,\eqref{Noncommute}). Hence overall the Matrix $M$ may not be Hermitian, in general, i.e. $(M^{\dagger}\neq M)$. For two qubits, the structure of operators $\{W(t).V\}$ can be expanded in the composite Hilbert space $H_{1}\otimes H_{2}$ with the dimension $(2^{2}\times 2^{2})$.

We analytically found the structure of the matrix $M$ whose trace is invariant under the action of the operator $\sigma_{y} \otimes \sigma_{y}$ in the space $H_{1}\otimes H_{2}$. This form of the matrix $M$ and its transpose are given by
\begin{equation}
M=\left(
\begin{array}{cccc}
 a & b & c & -a \\
 d & e & e & f \\
g & h & h & i \\
 j & k & l & -j \\
\end{array}
\right),\  M^{T}=\left(
\begin{array}{cccc}
 a & d & g & j \\
 b & e & h & k \\
c & e & h & l \\
 -a & f & i & -j \\
\end{array}
\right) \label{xt}
\end{equation}
with
\[
\big|{\Tr}[M]|=|{\Tr}[M^{T}]\big|\,.
\]
Here $(a,b,c,d,e,f,g,h,i,j,k,l)$ are complex numbers. The transpose of the matrix $M$, i.e. $M^{T}$, is also trace invariant under the action of the operator $\sigma_{y}\otimes \sigma_{y}$. Investigation of the matrix $M$ depends on many factors such as the Hamiltonian $H$ of the physical system, the nature of the scrambling operators $\{W(0),V\}$ and the density matrix $\rho$ of the state. Finding out the exact form of the constitutes $\{H,W(0),V,\rho\}$, which produces the structure of $M$
is mathematically difficult problem. We recall that the trace of $M$ is invariant under the action of $\sigma_{y}\otimes \sigma_{y}$, so we obtain the following relation
\begin{figure*}[htp]
  \centering
  \subfigure[Plot of QI scrambling $\langle C(t)\rangle_\rho $ vs. fidelity $f$ and ${\Im}\{Z\}$.]{\includegraphics[scale=0.6]{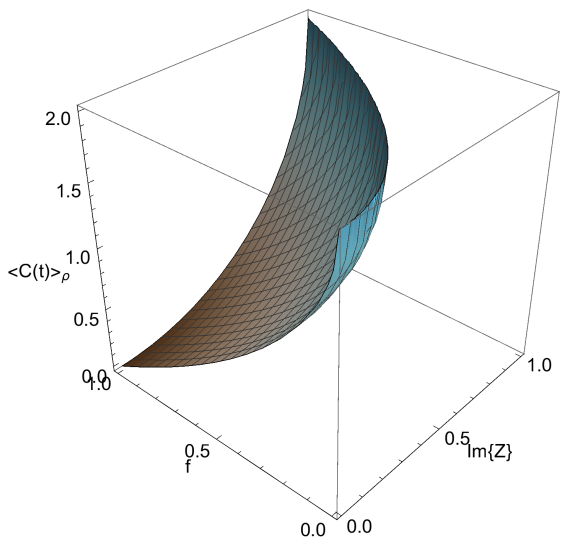}}\quad \quad
  \subfigure[Plot of Entanglement $C_{r}(M)$ vs. QI scrambling $\langle C(t)\rangle_\rho$ and ${\Im}\{Z\}$. ]{\includegraphics[scale=0.8]{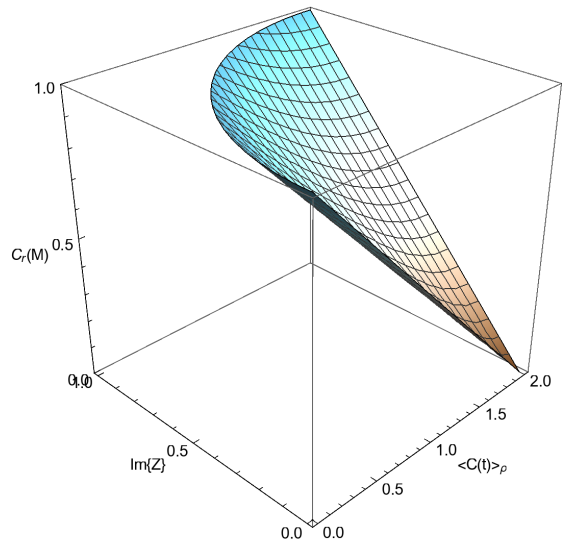}}
 \caption{Plot of $\langle C(t)\rangle_\rho$ and $C_{r}(M)$ with ${\Im}\{Z\}\neq 0$ for two-qubit states.} \label{cs2}
\end{figure*}
\begin{equation}
\big|{\Tr}[M]\big|=\big|{\Tr}[(\sigma_{y} \otimes \sigma_{y}).M]\big| \label{te1}
\end{equation}
or
\[
\big|{\Tr}[M^{T}]|=|{\Tr}[(\sigma_{y} \otimes \sigma_{y}).M^{T}]\big|\,.
\]
By substituting the value of $M$ from Eq.\,\eqref{d1}, we can write the Eq.\,\eqref{te1} as follows
\[
|{\Tr}[W(t).V.W(t).V.\rho]|=|{\Tr}[(\sigma_{y} \otimes \sigma_{y}).W(t).V.W(t).V.\rho]|
\]

Using the above relation, we can easily obtain the new form of Eq.\,\eqref{cc1}
\begin{equation}
C_{r}(M)=|{\Tr}[W(t).V.W(t).V.\rho]| \label{con1}
\end{equation}
Substituting the value of the factor $\sqrt{f}$ from Eq.\eqref{f1} in Eq.\eqref{con1}, we can establish the connection between concurrence and Uhlmann fidelity as given by
\begin{equation}
C_{r}(M)=\sqrt{f} \label{cf}
\end{equation}
Obtaining the value of $\sqrt{f}$ from the Eq.\eqref{bm1} and plug-in into Eq.\eqref{cf}, we can establish the direct connection between concurrence and Bures metric as
\[
C_{r}(M)=1-\frac{D^{2}}{2}\,.
\]
Putting the value of $f$ from Eq.\,\eqref{cf} in Eq.\,\eqref{es1}, we can obtain the direct connection between QI scrambling and concurrence as
\[
\langle C(t)\rangle_\rho=2\Big[1-\sqrt{\Big(C_{r}(M)\Big)^{2}-\Big({\Im}\{Z\}\Big)^{2}}\Big]\,. \label{cm1}
\]
As a consequence of the last equation, one can write the following expression for concurrence $C_{r}(M)$
\begin{equation}
C_{r}(M)=\sqrt{\Big[1-\frac{\langle C(t)\rangle_\rho}{2} \Big]^{2}+\Big[{\Im}\{Z\} \Big]^{2}}\,. \label{CrM}
\end{equation}
The obtained equation establishes the connection between QI scrambling and concurrence. It should be noted that because of~Eq.\,\eqref{cf} and $f\in [0,1]$ the expression occurring in Eq.\,\eqref{CrM} under the radical symbol satisfies the following inequality
\[
   0\leq \Big[1-\frac{\langle C(t)\rangle_\rho}{2} \Big]^{2}+\Big[{\Im}\{Z\} \Big]^{2} \leq 1\,.
\]
If ${\Im}\{Z\}=0$, then Eq.\,\eqref{CrM} is simplified to
\begin{equation}
C_{r}(M)=1-\Big[\frac{\langle C(t)\rangle_\rho}{2} \Big]\,. \label{qie}
\end{equation}
The above mathematical relations are helpful to study the direct influence of QI scrambling on two-qubit concurrence during the complete evolution with the following conditions:
\begin{itemize}
\item The density matrix $\rho$ deals with two-qubit pure quantum states having real elements.
\item The density matrix after the complete evolution has the form given by Eq.\,\eqref{xt}.
\end{itemize}

If the exact structure of $M$ given in Eq.\,\eqref{xt} is not obtained, then one is forced to study the concurrence by using the Eq.\,\eqref{conc1}. The Uhlmann fidelity and Bures metric, both are measures of the degree of mismatch of quantum state during forward and backward evolution. We recall that in the present work we only focus on Uhlamann fidelity. We divide our study in two cases as given below.

\subsection{Case 1: ${\Im}\{Z\}=0$}
In this subsection we discuss the behavior of QI scrambling $\langle C(t)\rangle_\rho$, concurrence $C_{r}(M)$ and fidelity $f$ when the Eq.\,\eqref{ImZ:0} holds in all the corresponding mathematical expressions.
We plot the QI scrambling and concurrence vs. the Uhlmann fidelity $f$ by using the Eqs.\,\eqref{qz1} and \eqref{cf} in the left part of Fig.\ref{fg1}. We find that the QI scrambling is monotonically decreasing and the entanglement expressed by concurrence is a monotonically increasing function. The decreasing nature of QI scrambling also comes from the series given in Eq.\,\eqref{se1}, as this series is a convergent one. Both the graphs intersect at $f\approx 0.44$, hence QI scrambling and entanglement both have equal value at this point. We refer to this point as QI scrambling and concurrence balancing point. At $f=1$, the QI scrambling becomes zero, which means that the operators $W(0),V$ commute and contribute in no scrambling but the concurrence sustains to $C_{r}(\rho)=1$.

The decreasing nature of QI scrambling leads to decay of the entanglement and hence may not be a favorable candidate in quantum information processing. This may also be observed in the right figure incorporated in Fig.\ref{fg1}, which is a plot of entanglement vs. QI scrambling by using the Eq.\,(\ref{qie}). We find, when QI scrambling is minimal, $\langle C(t)\rangle_\rho=0$, the entanglement is maximal, i.e. $C_{r}(\rho)=1$, but increasing QI scrambling decreases the entanglement linearly.

\subsection{Case 2: ${\Im}\{Z\}\neq 0$}

In this subsection we consider first the behavior of QI scrambling $\langle C(t)\rangle_\rho$ and fidelity $f$, with nonzero imaginary part of $Z$ dealing with Eq.\,\eqref{es1}. The graphical results are shown in Fig.\,\ref{cs2}\,(a). We find that the QI scrambling is zero at the point
\[ \{f=1,{\Im}\{Z\}=0\}
\]
and grows in its value when $f - {\Im}\{Z\}^2$, which is non-negative because of the equality \eqref{f1}, decreases. The scrambling achieves its maximum value $\langle C(t)\rangle_\rho=2$ at
\[
f - {\Im}\{Z\}^2=0\,.
\]
Then, in Fig.\ref{cs2}\,(b) we plot the action of QI scrambling and the factor $({\Im}\{Z\})$ on the entanglement characterized by the concurrence $C_{r}(M)$. This dependence is given by Eq.\,\eqref{CrM}. We establish the mere fact that as the amount of scrambling increases, it leads to decay of the entanglement in the system. Hence we find the destroying behavior of QI scrambling such that it may not serve as a useful factor for quantum processes based on entanglement.

\section{Dynamics of Quantifiers in Bell states}

In this section we study the dynamics of QI scrambling, Uhlmann fidelity and concurrence by considering two qubits prepared in Bell states
\begin{eqnarray}
&& |\psi_{\pm}\rangle_{1}=\frac{1}{\sqrt{2}}[|00\rangle\pm|11\rangle]\,\label{b1} \\
&& |\psi_{\pm}\rangle_{2}=\frac{1}{\sqrt{2}}[|01\rangle\pm|10\rangle]\,.\label{b2}
\end{eqnarray}

\begin{figure*}
\includegraphics[scale=1.3]{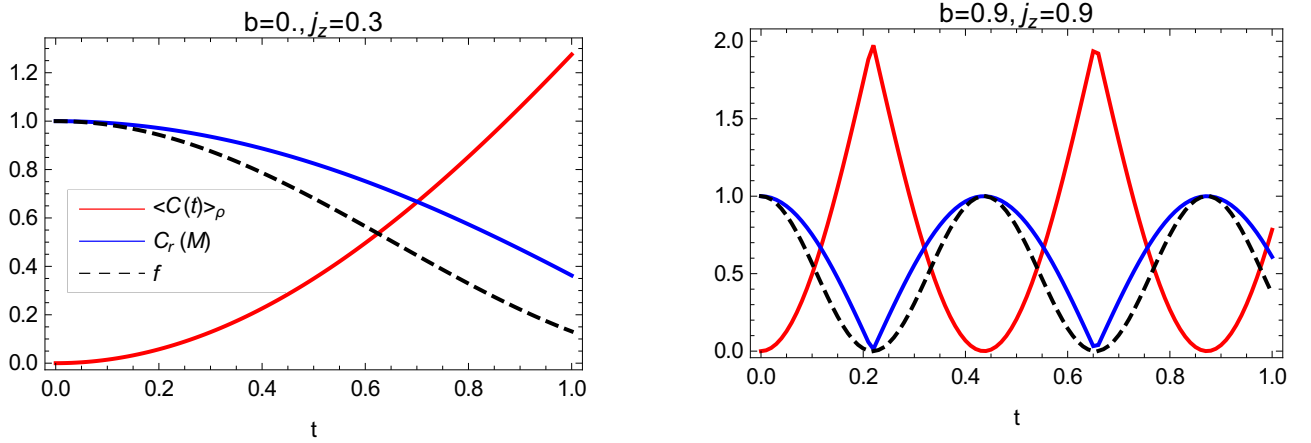}
\caption{Time evolution of QI scrambling $\langle C(t)\rangle_{\rho}$, fidelity $f$ and concurrence $C_{r}(M)$} for Bell states \label{bell1}
\end{figure*}
The corresponding density matrices of these states are expressed as $(\rho_{\pm})_{1}$ and $(\rho_{\pm})_{2}$. We assume that the two qubits are prepared in Bell states and carry the Ising interaction with external imposed magnetic filed in z direction. This Ising Hamiltonian is expressed as
\[
H=-j_{z}\sum_{i=1}^{2}\sigma^{z}_{1}\sigma^{z}_{2}-b\sum_{i=1}^{2}\sigma_{z}.
\]
The operator $W(0)$ evolves under the Hamiltonian $H$ and follows the series given in Eq.\,\eqref{se1}. This series can be easily derived by using the famous Baker-Campbell-Hausdorff formula\,\cite{bch1,bch2}. The rate of change of this series plays an important role in QI scrambling. To proceed the dynamical study of QI scrambling and concurrence, we consider here the scrambling operators as Pauli ones such that
\[
\{W(0),V \}=\{\sigma_{i},\sigma_{j}\}
\]
with
\[
\{\sigma_{i},\sigma_{j}\} \in \{\sigma_{x},\sigma_{y},\sigma_{z}\}, \quad (i,j)\in \{x,y,z\}\,.
\]
All the Pauli operators are  Hermitian as well as unitary matrices in agreement with the Eqs.\,\eqref{o1},\eqref{o2} and fulfill the need of choosing $\{W(0),V\}$. We consider the case when any one of the two qubits evolves under the action of Pauli operators. For example, let these operators act on the first qubit. In this direction we can develop the structure of the operators $\{W(0),V\}$ for two qubits in the composite Hilbert space $H_{1}\otimes H_{2}$ as given by
\begin{equation}
\{W(0),V\}_{(i,j)}=\{\sigma_{i}\otimes I,\sigma_{j}\otimes I\}\,. \label{w1}
\end{equation}
Here we mention that the factor
\[
{\Tr}[W(t).V.W(t).V.\rho]
\]
 is responsible for QI scrambling and concurrence obtained in Eqs.\,(\ref{es1}) and (\ref{con1}), respectively. This factor is invariant under the cyclic permutations of the operators. We consider the following option
\begin{equation}
\{W(0),V\}_{(j,i)}=\{\sigma_{j}\otimes I,\sigma_{i}\otimes I\}\,.\label{w2}
\end{equation}
Then, because of the cyclic permutation property, both Eqs.\eqref{w1} and \eqref{w2} will produce the same results of QI scrambling and entanglement. So we are led to choose the following combinations of the operators $\{W(t),V\}$ for the current study as
\begin{eqnarray*}
&\{\sigma_{x}\otimes I,\sigma_{x}\otimes I \}\,,\quad
&\{\sigma_{x}\otimes I,\sigma_{y}\otimes I \} \nonumber \\
&\{\sigma_{x}\otimes I,\sigma_{z}\otimes I \}\,,\quad
&\{\sigma_{y}\otimes I,\sigma_{y}\otimes I \} \nonumber \\
&\{\sigma_{y}\otimes I,\sigma_{z}\otimes I \}\,,\quad
&\{\sigma_{z}\otimes I,\sigma_{z}\otimes I \}\nonumber
\end{eqnarray*}
We study all the Bell states given in Eqs.\,\eqref{b1} and \eqref{b2} under the above mentioned combinations of operators.   We find that all the Bell states $(\rho_{\pm})_{1}$ and $(\rho_{\pm})_{2}$ under the actions of the combinations
$$\{\sigma_{x}\otimes I,\sigma_{x}\otimes I \},\{\sigma_{x}\otimes I,\sigma_{y}\otimes I\},\{\sigma_{y}\otimes I,\sigma_{y}\otimes I\}$$
produce the same results for all the quantifiers such that QI scrambling $\langle C(t)\rangle$, fidelity $f$, Bures metric and concurrence $C_{r}(M)$. The functions of these quantifiers are obtained as follows:
\begin{eqnarray*}
\text{QI scrambling:}\quad \langle C(t)\rangle=
2 \left(1-\cos[4 t (b+j_{z})]\right)\label{p1}\,, \\
\text{Uhlmann Fidelity:}\quad f=\cos^2[(4 t (b+j_{z}))]\,, \\
\text{Bures Metric:}\quad D=\sqrt{2}\cdot\sqrt{1-\cos[4 t (b+j_{z})]}\,, \\
\text{Concurrence:}\quad C_{r}(M)=\cos[4 t (b+j_{z})]\,.
\end{eqnarray*}
On the other hand the actions of the operators
$$\{\sigma_{x}\otimes I,\sigma_{z}\otimes I\},\{\sigma_{y}\otimes I,\sigma_{z}\otimes I\},\{\sigma_{z}\otimes I,\sigma_{z}\otimes I\}$$ do not produce any scrambling for all the states $(\rho_{\pm})_{1}$ and $(\rho_{\pm})_{2}$. Corresponding to these action operators, the values of all the quantifiers \{QI scrambling, Uhlmann Fidelity, Bures Metric and Concurrence\} are obtained as $\{0,1,0,1\}$. The functions occurring in all the quantifiers are oscillating functions with the parameters $(b,j_{z},t)$. Here we plot the dynamical behavior of the quantifiers QI scrambling, Uhlmann fidelity and concurrence with the varying parameters $(b,j_{z})$ vs. the parameter time $t$ in Fig.\,\ref{bell1}.

We find the peak value of the QI scrambling vanish the entanglement in the system and makes it zero. The QI scrambling and entanglement balancing points are periodic in Bell states. On the other hand we also do analysis of QI scrambling vs. Uhlmann fidelity. As the Uhlmann fidelity $f$ is zero, it means the degree of mismatch between forward and backward evolution of the quantum state $|\psi\rangle$ is very high and hence the QI scrambling is also very high, which consequently kills the entanglement in the system. If the Ulhmann fidelity $f$ approaches to the amplitude as unity, then the forward and backward evolution of the state $|\psi\rangle$ is the same and QI scrambling is zero in the states, which helps to keep the high amount of the entanglement in Bell states.

\section{Conclusion}

In this article we established the mathematical connections among the quantifiers for QI scrambling, Uhlmann fidelity, Bures metric and bipartite concurrence. Most importantly, a mathematical relation is developed between QI scrambling and concurrence, which is useful to study the direct influence of QI scrambling on bipartite entanglement for real density matrices.

Further, we have studied the dynamical behavior of quantum information scrambling, Uhlmann fidelity and entanglement with Ising Hamiltonian in two qubits prepared in Bell states. We have used the scrambling operators as Pauli operators and determined the combinations of scrambling operators under which no scrambling takes place in Bell states.

The influence of QI scrambling on entanglement has been studied and thereby established that the increasing amount of QI scrambling decreases the entanglement in the system and may not be a good candidate in quantum information processing. In addition, we found that the dynamics of all the quantifiers in Bell states is periodic. It is also investigated that the QI scrambling and entanglement balancing points are periodic as well. The present work may be explored in larger perspectives and may be helpful for the quantum information community.

\section{Acknowledgements}
This work was supported by the "RUDN University Program (5-100)". We thank the anonymous reviewer for valuable comments and remarks and also Daniel Robertz for improving the use of English in the manuscript

\end{document}